\documentclass[a4paper,12pt]{article}
\usepackage{epsfig}
\begin{document}

\large
\begin{center}

{\Large \bf OPTICAL SPECTROSCOPY OF MT DRA IN 2006 AND 2009}

\bigskip

A.M. Zubareva$^{1,2}$, V.V. Shimansky$^3$, N.V. Borisov$^4$, A.F.
Valeev$^4$

\end{center}

$^1$Institute of Astronomy, Russian Academy of Sciences, 48
Pyatnitskaya Str., Moscow 119017, Russia; zubareva@inasan.ru

$^2$Sternberg Astronomical Institute, Moscow University, 13
University Ave., Moscow 119992, Russia

$^3$Kazan Federal University, 18 Kremlevskaya Str., Kazan 420008,
Russia

$^4$Special Astrophysical Observatory, Russian Academy of
Sciences, Nizhny Arkhyz 369167, Russia

\bigskip

{\normalsize {\it ABSTRACT.} We present low-resolution
phase-resolved spectra of the polar MT Dra during its high states
in 2006 and 2009. Balmer series, He I and He II 4686 \AA\ emission
lines have complex shapes and show similar profile variations over
an orbital cycle. The radial velocities vary with orbital period
and display motions of the gas falling close to the magnetic poles
of the white dwarf.}

\bigskip

\section{INTRODUCTION}

AM Her type variables or polars are cataclysmic binaries which
contain a magnetic white dwarf with fields from $10^7$ to
$10^8$~G. The behaviour of the material flow from the red dwarf
secondary is totally controlled by the magnetic field. The stream
attached to the field lines is funneled to the white dwarf's
magnetic pole where an accretion column is formed. In the case of
two magnetic poles active, two accretion columns exist. The
location and shape of  the accretion column will depend on the
orientation of the magnetic dipole and on the location of the
coupling region (see Warner 1995 for a review).

MT Dra (RX J1846.9+5538) is one of several polars with accretion
onto two poles (Schwarz et al. 2002). During our sets of
observations, the system was in a high brightness state. Here we
present preliminary results of the analysis of our spectroscopic
data.

\section{OBSERVATIONS}

Our spectroscopy of MT Dra was performed in the Special
Astrophysical  Observatory of the Russian Academy of Sciences
(Nizhny Arkhyz, Russia) in two observing runs, on March 21--22,
2006 and October 21--22, 2009. Time-resolved spectra were obtained
with the 6-m telescope using the SCORPIO focal reducer (Afanasiev
and Moiseev 2005) in a long-slit mode, with a slit width of $1''$,
providing a spectral resolution of 5~\AA. The EEV-42-40
$2048\times2048$-pixel CCD camera and a grating with 1200 grooves
per mm together yield the dispersion of 0.88~\AA/pixel. On the two
occasions, we obtained 34 spectra over almost three orbital
cycles. The acquired spectra cover a wavelength range from
3900~\AA\ to 5700~\AA. Individual target exposure times were 300~s
for 16 spectra in 2006; 900~s for 2 spectra and 600~s for 16
spectra in 2009.

We performed data reductions using the standard MIDAS packages.
The procedure included standard bias, flat-field correction, and
cosmic particles removal procedures, and optimal extraction of
spectra, followed by wavelength calibration using He-Ne-Ar lamp
frames. Typical continuum signal-to-noise ratio of an individual
spectrum for the data of 2006 is $\sim30$ and for the data of
2009, $\sim40$. An example of the spectrum is given in Fig.~1.

\section{RESULTS}

During our observing runs, MT~Dra was in a high brightness state.
The spectra contain only emission lines and show no signs of the
secondary. They are typical of polars in a high brightness state.
Balmer series lines with a flat decrement, He~I and strong He~II
lines are present. The N~III${+}$C~III complex (Bowen blend) at
4640--4650~\AA\ is also seen. Parts of the normalised spectra containing
H$\beta$ and HeII~4686~\AA\ lines for both 2006 and 2009 are shown
in Fig.~2. We use the ephemeris HJD$_{\rm min} = 2454676.446 +
0.0893869\times E$ from Zubareva et al. (2011).

The line profiles have a very complex structure. They are
asymmetric and show two or more peaks at some phases.
Unfortunately, the resolution of the spectra is insufficient to
distinguish between individual line components.

The radial velocities of H$\beta$ and HeII~4686~\AA\ lines were
measured taking into account weights of every component of a line.
Semi-amplitudes of the curves for both runs are close to
600~km~s$^{-1}$ (Figs. 3 and 4). Also given in the figures are the
line widths (FWHM). The fact that radial velocities show almost
sinusoidal changes is quite unusual. Since MT Dra was in its
high-mass-transfer state during both observing runs, one would
expect both accretion columns to contribute to the emission of the
system. We suppose that the separation between the lines
corresponding to different accretion columns is quite small, so
the  centroid velocities follow a sine wave.

The exact inclination of the system and the orientation of the
magnetic dipole are still unknown. However, in some aspects MT Dra
is similar to the well-studied polar ST~LMi (Robertson et al.
2008) except that only one magnetic pole is active in ST~LMi most
of the time. The phase shift of $\sim 0.2$ between the phased
radial velocity curves for 2006 and 2009 runs could be explained
using geometric considerations. Due to precession of the magnetic
dipole axis around the rotation axis, we can see an accretion
column (not necessarily cylindrical) in different positions at
different phases (Ferrario and Wickramasinghe 1990).

Our observations were carried out under different conditions
during the two runs and thus do not allow us to draw a more
detailed picture. In order to get more information about MT~Dra on
the base of solely the optical region of the spectrum,
simultaneous spectroscopy and photometry are necessary.

\bigskip

{\it Acknowledgements.} AMZ would like to thank E.A.~Barsukova
(Special Astrophysical Observatory) and Yu.V.~Pakhomov (Institute
of Astronomy) for their help in processing the spectra.

\bigskip

REFERENCES

Afanasiev, V.L. and Moiseev, A.V. 2005, Astron. Let., {\bf 31},
194

Ferrario, L. and Wickramasinghe, D.T. 1990, Astropys. J., {\bf
357}, 582

Robertson, J.W., Howell, S.B., Honeycutt, R.K., et al. 2008,
Astron. J., {\bf 136}, 1857

Schwarz, R., Greiner, J., Tovmassian, G.H., et al. 2002, Astron.
and Astrophys., {\bf 392}, 505

Warner, B. 1995, Cataclysmic Variable Stars (Cambridge University
Press, Cambridge)

Zubareva, A.M., Pavlenko, E.P., Andreev, M.V., et al. 2011,
Astron. Rep, {\bf 55}, 224

\bigskip

\begin{figure}[!h]
\centerline{\hbox{\psfig{figure=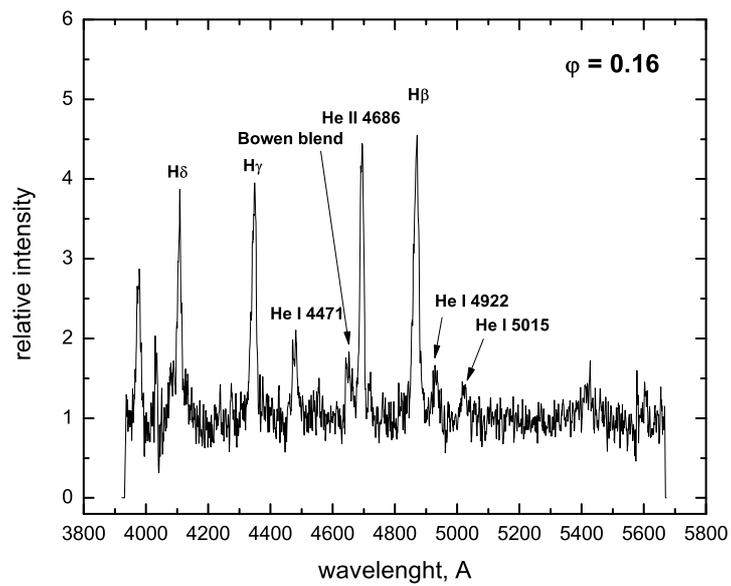,angle=0,clip=,width=12cm}}}
\caption[]{A spectrum from the 2006 data set. The orbital phase is
given.} 
\end{figure}

\begin{figure}[!h]
\centerline{\hbox{\psfig{figure=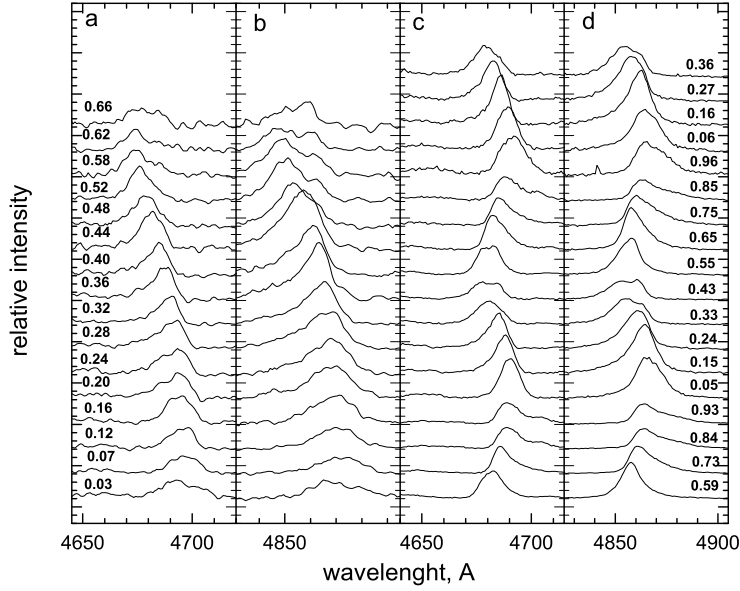,angle=0,clip=,width=12cm}}}
\caption[]{Trailed spectra for He II 4686~\AA\ line (a, c) and
H$\beta$ line (b, d) for both 2006 and 2009 data runs. The orbital
phases for 2006 data are indicated on panel a, for 2009 data -- on panel d.} 
\end{figure}

\begin{figure}[!h]
\centerline{\hbox{\psfig{figure=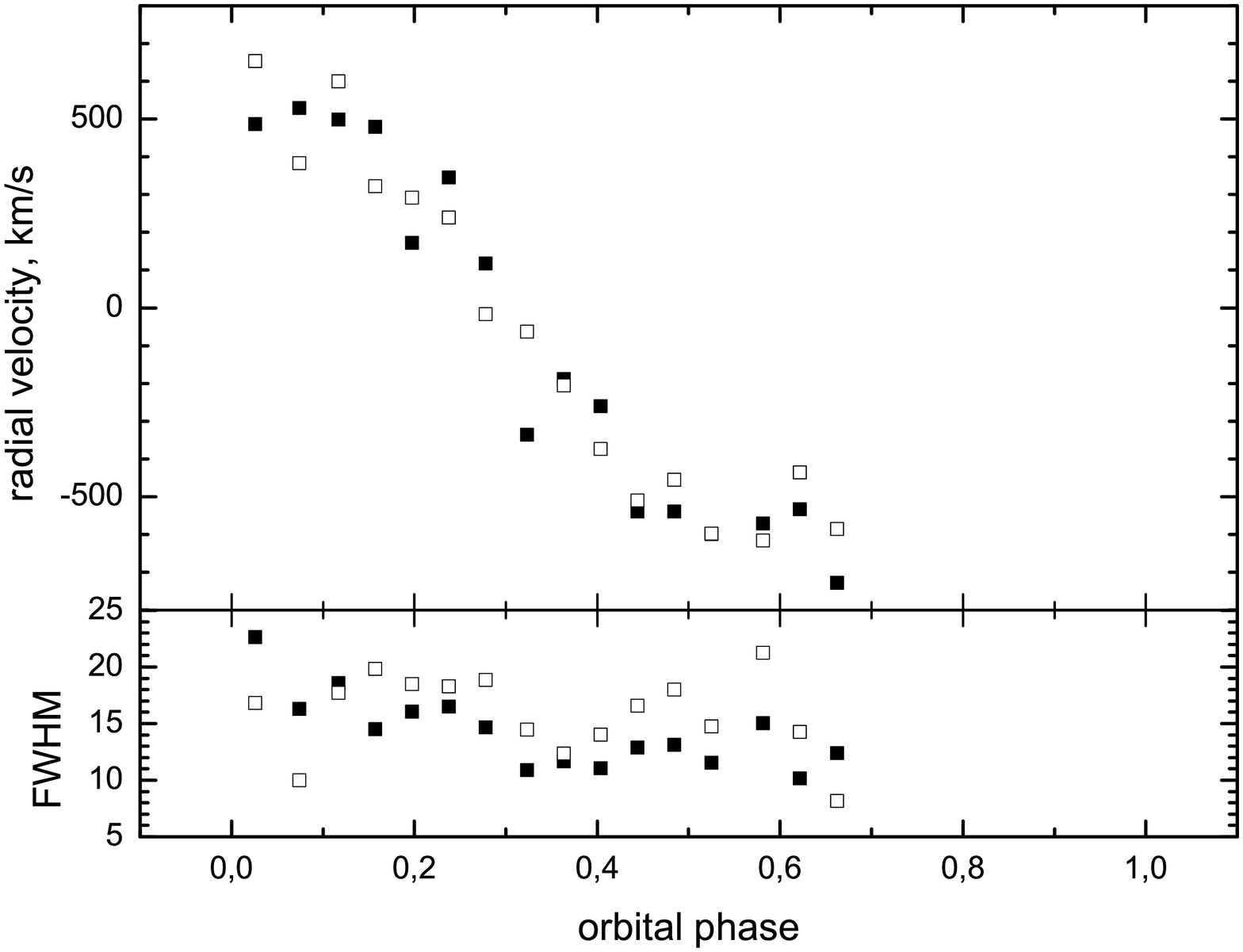,angle=0,clip=,width=12cm}}}
\caption[]{MT Dra radial velocities and line widths (FWHM) from
the 2006 data set. Open squares are for H$\beta$, filled squares
are for He II 4686~\AA.} 
\end{figure}

\begin{figure}[!h]
\centerline{\hbox{\psfig{figure=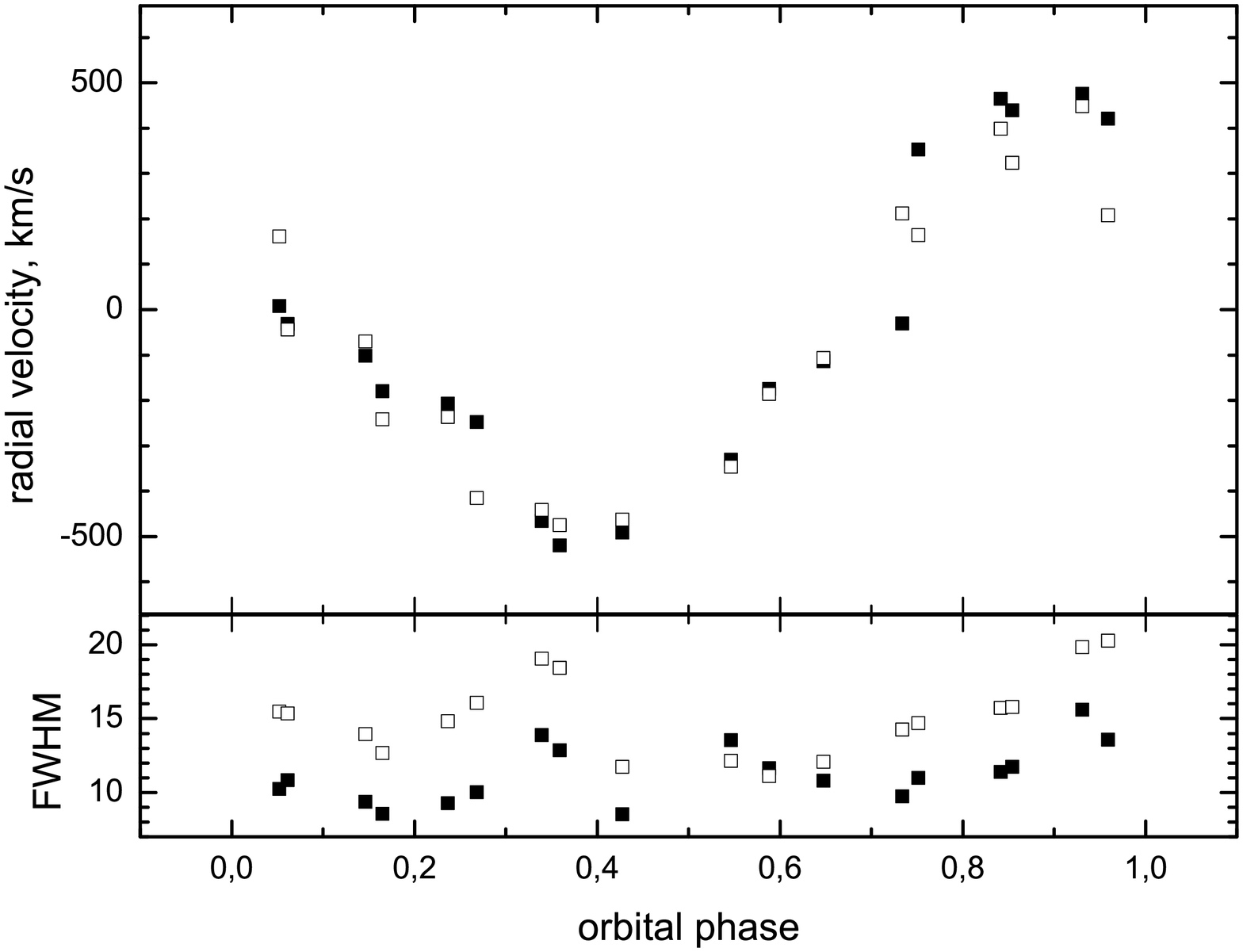,angle=0,clip=,width=12cm}}}
\caption[]{MT Dra radial velocities and line widths (FWHM) from
the 2009 data set. Open squares are for H$\beta$, filled squares
are for He II 4686 \AA.} 
\end{figure}

\end{document}